\documentclass[aps,showpacs,twocolumn]{revtex4}
\usepackage{amsfonts}
\usepackage{amsmath}
\usepackage{amssymb}
\usepackage{graphicx}
\usepackage{bm}

\input{tcilatex}

\begin{document}

\title{Comment on "Curvature capillary migration of microspheres" by N.
Sharifi-Mood, I.B. Liu, K.J. Stebe, Soft Matter, 2015, 11, 6768,  arXiv:1502.01672}
\author{Alois W\"{u}rger}
\affiliation{Laboratoire Ondes et Mati\`{e}re d'Aquitaine, Universit\'{e} de Bordeaux \&
CNRS, 33405 Talence, France}

\begin{abstract}
In a recent paper, Sharifi-Mood et al. study colloidal particles trapped at
a liquid interface with opposite principal curvatures $c_{1}=-c_{2}$. In the
theory part, they claim that the trapping energy vanishes at second order in 
$\Delta c=c_{1}-c_{2}$, which would invalidate our previous result [Phys. Rev. E, 2006, 74, 041402]. Here we show that this claim arises from
an improper treatment of the outer boundary condition on the deformation
field. For both pinned and moving contact lines, we find that the outer
boundary is irrelevant, which confirms our previous work. More generally, we
show that the trapping energy is determined by the deformation close to the
particle and does not depend on the far-field. 
\end{abstract}

\maketitle

Colloidal particles trapped at a curved liquid interface are subject to
capillary forces that do not depend on their mass or charge but on
geometrical parameters only. In Ref. \cite{Sha15}, Sharifi-Mood et al.
provide an interesting analysis of the role of contact line pinning.
Regarding the trapping energy, however, these authors assert that it
vanishes at second order, contrary to previous work, and they state that\
\textquotedblleft the origin of the discrepancy is an inappropriate
treatment of the contour integral\textquotedblright\ in \cite{Wue06}. The
present comment intends to refute this claim of \cite{Sha15}, to
unambiguously determine the trapping energy, and to clarify the role of the
far-field.

Previous works \cite{Wue06,Lea13,Bla13} rely on the assumption that\
curvature-induced forces arise from the interface close to the particle and
that the far-field is irrelevant. Thus the profile of the bare interface is
taken in small-gradient approximation, $h_{0}=\frac{\Delta c}{4}\cos
(2\varphi )r^{2}$, which is valid only at distances shorter than the
curvature radius $R_{c}=1/\Delta c$. Adding a particle modifies the profile
as $h=h_{0}+\eta $, where the deformation field 
\begin{equation}
\eta =\frac{\Delta c}{12}\cos 2\varphi \frac{r_{0}^{4}}{r^{2}}.  \label{2}
\end{equation}%
is determined from the contact angle at the particle surface. By the same
token, the trapping energy comprises only near-field contributions, and is
given by the boundary term along the contact line of radius $r_{0}$, 
\begin{equation}
E_{\text{in}}=\gamma \oint\nolimits_{r_{0}}\eta \mathbf{\nabla }h_{0}\cdot 
\mathbf{n}ds=-\frac{\pi }{24}\gamma r_{0}^{4}\Delta c^{2}.  \label{6}
\end{equation}%
Since a similar line integral of $\eta \mathbf{\nabla }\eta $ is cancelled
by the area change due to displacement of the contact line on the particle
surface, one obtains the curvature-dependent energy $E=E_{\text{in}}$ \cite%
{Wue06}, which was confirmed in \cite{Lea13,Bla13}.

Sharifi-Mood et al. attempt to go beyond the near-field approach and to
evaluate the term arising at the outer boundary, 
\begin{equation}
E_{\text{out}}=\gamma \oint\nolimits_{R_{\text{out}}}\eta \mathbf{\nabla }%
h_{0}\cdot \mathbf{n}ds,  \label{7}
\end{equation}%
where, in the simplest geometry, $R_{\text{out}}(\varphi )$ is the distance
of the container wall from the particle. Using the above expressions $%
h_{0}\propto r^{2}$ and $\eta \propto r^{-2}$, and letting $R_{\text{out}%
}\rightarrow \infty $, these authors find in (24) of \cite{Sha15} the
relation $E_{\text{out}}=-E_{\text{in}}$. This leads them to the conclusion
that the trapping energy vanishes at second order, $E=E_{\text{in}}+E_{\text{%
out}}=O(\Delta c^{4})$.

Yet this argument is flawed by the fact that $E_{\text{out}}$ is calculated
with the near-field deformation (\ref{2}) which is not correct at $R_{\text{%
out}}$. (Moreover, $h_{0}\propto r^{2}$ is valid at distances within the
curvature radius only \cite{Gal15}.) In the following we evaluate $E_{\text{%
out}}$ with the the correct deformation field $\eta $, which satisfies
appropriate conditions at the outer boundary. For both pinned and moving
contact lines, we find that the outer boundary does not contribute to $E$,
and thus confirm the results from the near-field approach and in particular
the trapping energy $E=E_{\text{in}}$ of \cite{Wue06}.

\begin{figure}[ptb]
\includegraphics[width=\columnwidth]{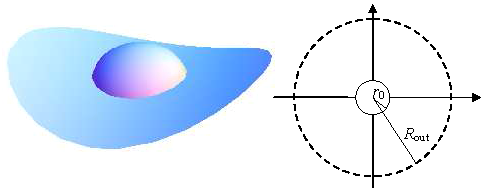}
\caption{Schematic view of a particle trapped at a liquid
interface with zero mean curvature, $c_{1}+c_{2}=0$. The right panel shows
the simple case of a circular interface of radius $R_{\text{out}}$ (dashed
line); the inner circle indicates the contact line of radius $r_{0}$ on the
particle.}
\end{figure}

\section*{Pinned outer contact line}

In many experiments, the outer boundary pins the interface at a fixed
contact line, $h_{0}|_{R_{\text{out}}}=K(\varphi )$; examples are the
micropost in Fig. 5 of \cite{Sha15} or pinning due to the surface roughness
of the container wall.\ The same contact line delimits the deformed profile, 
$h|_{R_{\text{out}}}=K(\varphi )$, which implies 
\begin{equation}
\eta |_{R_{\text{out}}}=(h-h_{0})|_{R_{\text{out}}}=0.  \label{9}
\end{equation}%
As a consequence, the contour integral in (\ref{7}) vanishes, and the
trapping energy is $E=E_{\text{in}}$.\ 

As a simple example, we consider a particle at the center of a circular
interface with constant $R_{\text{out}}$, as in Fig. 1. One readily finds
the form 
\begin{equation}
\eta =\frac{\Delta c}{12}\cos 2\varphi \left( \frac{r_{0}^{4}}{r^{2}}-\frac{%
r_{0}^{4}r^{2}}{R_{\text{out}}^{4}}\right) \xi ,  \label{8}
\end{equation}%
which solves Laplace's equation $\nabla ^{2}\eta =0$ and satisfies (\ref{9})
and thus $E_{\text{out}}=0$. (The factor $\xi =1/(1+r_{0}^{4}/R_{\text{out}%
}^{4})\approx 1$ assures the boundary condition on the particle surface.)

\section*{Moving contact line}

Though the preceding paragraphs refute the claim of \cite{Sha15}, we
complete the discussion by considering the case where the contact line at
the outer boundary is not pinned but moves on the confining walls. Then the
boundary condition involves the gradient of the profile, $L(\varphi )=%
\mathbf{n\cdot \nabla }h_{0}|_{R_{\text{out}}}$. The same condition applies
to the deformed interface $h=h_{0}+\eta $, implying 
\begin{equation}
\mathbf{n\cdot \nabla }\eta |_{R_{\text{out}}}=\mathbf{n\cdot \nabla }%
(h-h_{0})|_{R_{\text{out}}}=0.  \label{10}
\end{equation}%
It is straightforward to show that the corresponding boundary term $E_{\text{%
out}}$ does not vanish, yet is cancelled by the area change $E_{A}$ due to
the moving contact line, resulting in $E=E_{\text{in}}$. (In previous work,
the area change has been considered at the inner boundary only; cf. $E_{P}$
in \cite{Wue06} or Eq. (26) in \cite{Sha15}.) For the circular interface of
Fig. 1, the deformation field is\ similar to (\ref{8}), albeit with a plus
sign instead of the minus; one readily calculates the integrated term at the
outer boundary $E_{\text{out}}=-2E_{\text{in}}$, and the area change $E_{%
\text{A}}=2E_{\text{in}}$. Since these terms cancel each other, one has $%
E=E_{\text{in}}$.

In summary, when properly treating the outer boundary condition, for both
pinned and moving contact lines, we find as a rather general result that the
outer boundary does not contribute to the trapping energy. This invalidates
the claim $E=O(\Delta c^{4})$ of \cite{Sha15}, supports the near-field
approach of \cite{Wue06}, and confirms the trapping energy $E=E_{\text{in}}$
obtained previously. Retaining this quadratic term would modify the analysis
of the experimental results of \cite{Sha15}, for example the fits in Fig.\
7, yet does not affect the qualitative interpretation in terms of contact
line pinning.

\end{document}